\RequirePackage{ifpdf}
\ifpdf 
\documentclass[pdftex]{sigma}
\else
\documentclass{sigma}
\fi

\def\CL {{\cal L}}

\numberwithin{equation}{section}

\begin{document}

\newcommand{\mf}[1]{\mathfrak{#1} }
\newcommand{\mc}[1]{\mathcal{#1}}
\newcommand{\partiell}[2]{\frac{\partial#1 }{\partial#2 }}
\newcommand{\partl}[1]{\frac{\partial}{\partial#1}}
\newcommand{\funktional}[2]{\frac{\delta#1 }{\delta#2 }}
\newcommand{\de}{{\bf d}\!}
\newcommand{\To}{\rightarrow}
\newcommand{\abs}[1]{\mid#1 \mid}
\newcommand{\bei}[2]{\left. #1 \right| _{#2 }}
\newcommand{\dann}{\Rightarrow}
\newcommand{\hoch}[1]{^{#1 }}
\newcommand{\tief}[1]{_{#1 }}
\renewcommand{\hoch}[1]{{}^{#1}}\renewcommand{\tief}[1]{{}_{#1}}\newcommand{\lqn}[1]{\lefteqn{#1}}
\newcommand{\os}[2]{\overset{\lefteqn{{\scriptstyle #1}}}{#2}}
\newcommand{\us}[2]{\underset{\lqn{{\scriptstyle #2}}}{#1}}
\newcommand{\ous}[3]{\underset{\lefteqn{{\scriptstyle #3}}}{\os{#1}{#2}}}
\newcommand{\zwek}[2]{\begin{array}{l}
 #1\\
#2\end{array}}
\newcommand{\drek}[3]{\begin{array}{c}
 #1\\
#2\\
#3\end{array}}
\newcommand{\one}{1\!\!1}
\newcommand{\fussend}{\diamond}
\newcommand{\eps}{\varepsilon}
\newcommand{\lamlam}{\lambda\cdots\lambda}
\newcommand{\roro}{\rho\ldots\rho}
\newcommand{\proj}{\Pi}
\newcommand{\propinv}{R}
\newcommand{\prop}{P}
\newcommand{\la}{\langle}
\newcommand{\ra}{\rangle}
\newcommand{\nn}{\nonumber}
\newcommand{\ddxmu}{\frac{\partial}{\partial x^\mu}}
\newcommand{\spa}{\ \ \ }
\newcommand{\tr}{\mathop{{\rm Tr}}}
\newcommand{\bo}{\mathbf}
\newcommand{\conus}{\cos (\omega t)}
\newcommand{\sinus}{\sin (\omega t)}
\newcommand{\str}{\mathop{{\rm Str}}}
\newcommand{\sn}{\mathop{{\rm sn}}}
\newcommand{\bt}{\beta}
\newcommand{\rem}[1]{}

\allowdisplaybreaks

\renewcommand{\PaperNumber}{061}

\FirstPageHeading

\ShortArticleName{Schwinger--Fronsdal Theory of
Abelian Tensor Gauge Fields}

\ArticleName{Schwinger--Fronsdal Theory\\  of
Abelian Tensor Gauge Fields}

\Author{Sebastian GUTTENBERG and George  SAVVIDY}

\AuthorNameForHeading{S.~Guttenberg and G.~Savvidy}

\Address{Institute of Nuclear Physics, Demokritos National Research Center,\\
Agia Paraskevi, GR-15310 Athens, Greece}

\Email{\href{mailto:guttenb@inp.demokritos.gr}{guttenb@inp.demokritos.gr}, \href{mailto:savvidy@inp.demokritos.gr}{savvidy@inp.demokritos.gr}}

\URLaddress{\url{http://hep.itp.tuwien.ac.at/~basti/},\url{http://www.inp.demokritos.gr/~savvidy/}$\!$}

\ArticleDates{Received April 23, 2008, in f\/inal form September 01,
2008; Published online September 04, 2008}

\Abstract{This review is devoted to the Schwinger and Fronsdal theory of
Abelian tensor gauge f\/ields. The theory describes the propagation of free massless gauge bosons
of integer helicities and their interaction with external currents.
Self-consistency of its equations requires only the traceless part of the
current divergence to vanish. The essence of the theory is given by the fact that
this weaker current conservation is enough to guarantee the unita\-ri\-ty of the theory.
Physically this means that only waves with
transverse polarizations are propagating very far from the sources.
The question whether such currents exist should be answered by a fully interacting theory.
We also suggest an equivalent representation of the corresponding action.}


\Classification{81T10; 81T13; 70S05; 70S10; 70S15; 35L05; 35L10}

\section{Introduction}

We shall start from the formulation of the  Schwinger--Fronsdal action
for symmetric Abelian tensor gauge f\/ield of rank $s$, $A_{\lambda_1 \dots \lambda_s}$
\cite{schwinger,fronsdal,singh}. The development which leads to the discovery of this
action and the corresponding review articles can be found in the extended literature
\cite{chang,van Dam:1970vg,Curtright:1979uz,Bengtsson:1983pd,Bengtsson:1983pg,
Bengtsson:1986kh,berends,Berends:1984wp,de Wit:1979pe,Bengtsson:2006pw,Savvidy:2006qf,
Bengtsson:2004cd,Francia:2002pt,Francia:2007qt,Sorokin:2004ie}.
The theory is gauge invariant, but to our  best knowledge
there is no unique and systematic way to extend this action to an interacting theory
from some sort of gauge principle.
This is in contrast with the Yang--Mills theory, where one can
formulate the gauge principle, to derive transformation properties of
the vector gauge f\/ield and to f\/ind out the corresponding gauge invariant action.

Therefore we shall postulate  the quadratic form for the Lagrangian $\CL
$
and then describe its invariant  and physical properties.
The variation of the Schwinger--Fronsdal action allows to derive the equation of
motion for a symmetric Abelian tensor gauge f\/ield of rank $s$, $A_{\lambda_1 \dots \lambda_s}$,
in the presence of an external current $J_{\lambda_1 \dots\lambda_s}$
\[
(LA)_{\lambda_{1}\ldots\lambda_{s}} =
J_{\lambda_{1}\ldots\lambda_{s}},
\]
where $L$ is a linear dif\/ferential operator of second order. As we shall see, the equation describes the
propagation of transverse polarizations of a spin-$s$ gauge boson and its interaction with the external current.
Self-consistency of this equation  requires that the traceless
part\footnote{Fields and currents are double traceless,
see Section~\ref{sec2} and especially footnote~\ref{fn:dbl-tl-current}.} of
the current divergence should vanish \cite{schwinger,fronsdal}
\[
\partial^{\mu}J_{\mu\lambda_{2}\ldots\lambda_{s}}-
\tfrac{1}{d+2s-6}\sum_{2}\eta_{\lambda_{2}\lambda_{3}}\partial^{\mu}J'_{\mu\lambda_{4}
\ldots\lambda_{s}}=0 .
\]
This is a weaker conservation law of the current, if one compares it with the
fully conserved current
$
\partial^{\mu}J_{\mu\lambda_{2}\ldots\lambda_{s}} =0.
$
{\it The weaker current conservation law
nevertheless guarantees the unitarity of the theory}~\cite{schwinger,fronsdal}.
Physically this means that
only waves with transverse polarizations are propagating very far from the sources,
as it is the case for fully conserved currents~\cite{feynman}.
It is outside of the  scope of this theory
to answer the question if such currents exist or not. It
should be answered by a fully interacting theory.
At the end of this review we shall also suggest an equivalent representation
of the corresponding action.

The subject which we do not touch in this review is the question
of possible extension of this
theory to a fully interacting theory. The answer still remains uncertain,
but self-consistency and beauty
of this theory tell us that probably some part of it  may become
essential in the construction of an interacting theory
\cite{berends,Berends:1984wp}.

For the recent development of interacting gauge f\/ield theories
based on the extension of the gauge principle
to {\it non-Abelian tensor gauge fields} see references
\cite{Savvidy:2005fi,Savvidy:2005zm,Savvidy:2005ki,Barrett:2007nn}
and for the calculation of the production cross section of spin-two
non-Abelian tensor gauge bosons see \cite{Konitopoulos:2008bd}.
The interacting f\/ield theories in anti-de
Sitter space-time background are reviewed in
\cite{Bekaert:2005vh,Vasiliev:1996,Vasiliev:1999ba,Engquist:2002vr,Sezgin:2001zs}.

\section[Schwinger-Fronsdal action]{Schwinger--Fronsdal action}\label{sec2}

The Schwinger--Fronsdal action for symmetric  Abelian tensor gauge f\/ields
of rank s was derived f\/irst for the rank-3 gauge f\/ield by Schwinger in \cite{schwinger}
and then was generalized by Fronsdal \cite{fronsdal} to arbitrary symmetric rank-$s$
f\/ield exploring the massless limit  of the Singh--Hagen action for massive tensor
f\/ields \cite{singh}. The massless action has the following form
\begin{gather}
S[A]  =  \int dx^{d}\;\tfrac{1}{2}\partial_{\mu}A_{\lambda_{1}
\ldots\lambda_{s}}\partial^{\mu}A^{\lambda_{1}
\ldots\lambda_{s}}-\tfrac{s}{2}\partial_{\mu}A^{\mu\lambda_{2}
\ldots\lambda_{s}}\partial^{\nu}A_{\nu\lambda_{2}\ldots\lambda_{s}}\nonumber \\
 \phantom{S[A]  =}{}  -\tfrac{s(s-1)}{2}A'_{\lambda_{3}
 \ldots\lambda_{s}}\partial_{\mu}\partial_{\nu}A^{\mu\nu\lambda_{3}
 \ldots\lambda_{s}}-\tfrac{s(s-1)}{4}\partial_{\mu}A'\tief{\lambda_{3}
 \ldots\lambda_{s}}\partial^{\mu}A'\hoch{\lambda_{3}\ldots\lambda_{s}} \nonumber\\
\phantom{S[A]  =}{} -\tfrac{s(s-1)(s-2)}{8}\partial_{\mu}A'^{\mu\lambda_{4}\ldots\lambda_{s}}\partial^{\nu}A'
 \tief{\nu\lambda_{4}\ldots\lambda_{s}}-A_{\lambda_{1}\ldots\lambda_{s}}J^{\lambda_{1}
 \ldots\lambda_{s}},\label{FronsdalAction}
 \end{gather}
where $A^{\lambda_{1}\ldots\lambda_{s}}$ is a symmetric Abelian tensor gauge f\/ield
of rank $s$ and $J^{\lambda_{1}\ldots\lambda_{s}}$ is a symmetric external  current. $A'$ denotes the trace of the f\/ield
$
A'\tief{\lambda_{3}\ldots\lambda_{s}}   \equiv   A^{\rho}\tief{\rho\lambda_{3}
\ldots\lambda_{s}}
$, while the other notations here should be self-evident.
The f\/ield is restricted to be double traceless\footnote{It was demonstrated by
Fierz and Pauli \cite{fierz} that in order to have a Lagrangian  description
of a spin-$s$ boson, one should introduce a traceless rank-$s$ tensor f\/ield together with
auxiliary traceless f\/ields of all lower ranks. Considering the massless limit of the
Singh and Hagen Lagrangian \cite{singh} one can prove that the tensors of rank $(s-3)$
and lower decouple and the remaining rank-$s$ and rank-$(s-2)$ tensors can be combined into a
single double-traceless f\/ield of rank~$s$~\cite{fronsdal}. Note that there exist
unconstrained formulations of the theory, with or without auxiliary f\/ields, which remove
the double-traceless constraint \cite{Francia:2002pt,Francia:2002aa,Bengtsson:2006pw,
Bengtsson:2004cd}, but lead to higher derivative or non-local terms.}, i.e.
\begin{gather}
A''_{\lambda_{5}\ldots\lambda_{s}}  \equiv
\eta^{\rho_{1}\rho_{2}}\eta^{\rho_{3}\rho_{4}}A_{\rho_{1}
\rho_{2}\rho_{3}\rho_{4}\lambda_{5}\ldots\lambda_{s}}=0.\label{hDoubleTraceless}
\end{gather}
The same property is inherited by  the current $J^{\lambda_{1}\ldots\lambda_{s}}$,
because it is contracted with the f\/ield $A_{\lambda_{1}\ldots\lambda_{s}}$ in the action,
thus $J''_{\lambda_{5}\ldots\lambda_{s}}=0$.
These conditions have an ef\/fect only for $s \geq 4$.

For $s=0$  the above action corresponds to a
massless scalar f\/ield interacting with an external current.
For $s=1$ only the f\/irst two terms contribute and
correspond to electrodynamics, and for $s=2$ one obtains
linearized gravity
\begin{gather}
s=0:\quad S  =  \int dx^{d}\;\tfrac{1}{2}\partial_{\mu}A\partial^{\mu}A - AJ, \nonumber\\
s=1:\quad S  =  \int dx^{d}\;\tfrac{1}{2}\partial_{\mu}A_{\lambda_{1}}\partial^{\mu}
A^{\lambda_{1}}-\tfrac{1}{2}\partial_{\mu}A^{\mu}\partial^{\nu}A_{\nu}-A_{\mu}J^{\mu}, \nonumber\\
s=2:\quad S  =  \int dx^{d}\;\tfrac{1}{2}\partial_{\mu}A_{\lambda_{1}\lambda_{2}}
\partial^{\mu}A^{\lambda_{1}\lambda_{2}}-\partial_{\mu}A^{\mu\lambda_{2}}\partial^{\nu}
A_{\nu\lambda_{2}}\nonumber \\
\phantom{s=2:\quad S  = }{} -A'\partial_{\mu}\partial_{\nu}A^{\mu\nu}-\tfrac{1}{2}\partial_{\mu}
 A'\partial^{\mu}A' -A_{\lambda_{1}\lambda_{2} }
 J^{\lambda_{1}\lambda_{2} } .\label{electromaggaravity}
\end{gather}
For $s=3$ it is the Schwinger action and has the following form \cite{schwinger}
\begin{gather*}
s=3:\quad S  =  \int dx^{d}\;\tfrac{1}{2}\partial_{\mu}
A_{\lambda_{1}\lambda_{2}\lambda_{3}}\partial^{\mu}
A^{\lambda_{1}\lambda_{2}\lambda_{3}}-\tfrac{3}{2}\partial_{\mu}
A^{\mu\lambda_{2}\lambda_{3}}\partial^{\nu}A_{\nu\lambda_{2}\lambda_{3}}\nonumber \\
\phantom{s=3:\quad S  =}{}  -3A'_{\lambda_{3}\ldots\lambda_{s}}\partial_{\mu}\partial_{\nu}
 A^{\mu\nu\lambda_{3}\ldots\lambda_{s}}-\tfrac{3}{2}\partial_{\mu}
 A'\tief{\lambda_{3}\ldots\lambda_{s}}\partial^{\mu}A'\hoch{\lambda_{3}\ldots\lambda_{s}} \nonumber \\
\phantom{s=3:\quad S  =}{} -\tfrac{3}{4}\partial_{\mu}A'^{\mu}\partial^{\nu}
 A'\tief{\nu}-A_{\lambda_{1}\lambda_{2}\lambda_{3}}
 J^{\lambda_{1}\lambda_{2}\lambda_{s}} .
\end{gather*}
Finally the Fronsdal action for $s=4$ is
\begin{gather*}
s=4:\quad S  =  \int dx^{d}\;\tfrac{1}{2}\partial_{\mu}A_{\lambda_{1}\ldots\lambda_{4}}
\partial^{\mu}A^{\lambda_{1}\ldots\lambda_{4}}-2\partial_{\mu}A^{\mu\lambda_{2}\lambda_{3}\lambda_{4}}\partial^{\nu}
A_{\nu\lambda_{2}\lambda_{3}\lambda_{4}}\nonumber \\
\phantom{s=4:\quad S  = }{}  -6A'_{\lambda_{3}\lambda_{4}}\partial_{\mu}\partial_{\nu}A^{\mu\nu\lambda_{3}\lambda_{4}}-3\partial_{\mu}
 A'\tief{\lambda_{3}\lambda_{4}}\partial^{\mu}A'\hoch{\lambda_{3}\lambda_{4}}\nonumber \\
\phantom{s=4:\quad S  = }{}   -3\partial_{\mu}A'^{\mu\lambda_{4}}\partial^{\nu}
 A'\tief{\nu\lambda_{4}}-A_{\lambda_{1}\ldots\lambda_{4}}
 J^{\lambda_{1}\ldots\lambda_{4}}.
 \end{gather*}
As we shall see later, the action (\ref{FronsdalAction}) is gauge invariant with respect to the
Abelian gauge transformation
\begin{gather}\label{generalgaugetransfromation}
\delta_{\xi}A_{\lambda_{1}\ldots\lambda_{s}} \equiv
\sum_{1}\partial_{\lambda_{1}}\xi_{\lambda_{2}\ldots\lambda_{s}},
\qquad\xi'_{\lambda_{4}\ldots\lambda_{s}}=0,
\end{gather}
where $\xi_{\lambda_{1}\ldots\lambda_{s-1}}$ is a symmetric gauge parameter  of rank $s - 1$
and the sum $\sum_{1}$ is over all inequivalent index permutations. The gauge parameter has to be
traceless, $\xi' =0$, as indicated. With such a restriction on the gauge parameter the class of
double-traceless f\/ields $\{A:A^{''}=0\}$  remains intact in the
course of gauge transformations. Indeed, the double trace of the f\/ield
transformation (\ref{generalgaugetransfromation}) is proportional to the trace of the
gauge parameter and therefore $\xi'$ should vanish. On the other hand, the
variation of the action with respect to the transformation
(\ref{generalgaugetransfromation}) is also proportional to $\xi'$ and vanishes only if $\xi'=0$.
We shall see this below. Because the gauge parameter $\xi$ is restricted to be
traceless, the corresponding symmetry group (\ref{generalgaugetransfromation}) is smaller and
as a result the
current conservation law is weaker (\ref{modified-current-cons}).
It seems that this may endanger the unitarity of the theory and our main concern is to demonstrate,
following Schwinger and Fronsdal \cite{schwinger,fronsdal}, that the theory is nevertheless unitary.
{\it Thus even with a smaller symmetry gauge group the theory still stays unitary!}

Let us derive the equation of motion. The variation of the action (\ref{FronsdalAction}) reads
\begin{gather}
\delta S  =  \int \delta A^{\lambda_{1}\ldots\lambda_{s}}
\Bigl\{-\partial^2 A_{\lambda_{1}\ldots\lambda_{s}}+s\partial_{\lambda_{1}}
\partial^{\nu}A_{\nu\lambda_{2}\ldots\lambda_{s}}-\tfrac{s(s-1)}{2}
\partial_{\lambda_{1}}\partial_{\lambda_{2}}A'_{\lambda_{3}\ldots\lambda_{s}}+\nonumber \\
\phantom{\delta S  =}{} -\tfrac{s(s-1)}{2}\eta_{\lambda_{1}\lambda_{2}}
 \Bigl(\partial^{\mu}\partial^{\nu}A_{\mu\nu\lambda_{3}\ldots\lambda_{s}}-
 \partial^2 A'_{\lambda_{3}\ldots\lambda_{s}}-\tfrac{s-2}{2}\partial_{\lambda_{3}}
 \partial^{\mu}A'_{\mu\lambda_{4}\ldots\lambda_{s}}\Bigr)-J_{\lambda_{1}
 \ldots\lambda_{s}}\Bigr\}.\label{FronsdalVar}
 \end{gather}
The variation of $A$ is restricted to be symmetric and double-traceless,
therefore the variational derivative $\funktional{S}{A}$ is equal to
the symmetric and double-traceless part of the terms in the curly bracket.
Let us f\/irst symmetrize the indices in the curly bracket. This yields
\begin{gather}
 -\partial^2 A_{\lambda_{1}
\ldots\lambda_{s}}+\sum_{1}\partial_{\lambda_{1}}\partial^{\nu}A_{\nu\lambda_{2}
\ldots\lambda_{s}}-\sum_{2}\partial_{\lambda_{1}}\partial_{\lambda_{2}}A'_{\lambda_{3}
\ldots\lambda_{s}}\nonumber \\
\qquad   -\sum_{2}\eta_{\lambda_{1}\lambda_{2}}\Bigl(\partial^{\mu}\partial^{\nu}
 A_{\mu\nu\lambda_{3}\ldots\lambda_{s}}-\partial^2 A'_{\lambda_{3}\ldots\lambda_{s}}-
 \tfrac{1}{2}\sum_{1}\partial_{\lambda_{3}}\partial^{\mu}A'_{\mu\lambda_{4}
 \ldots\lambda_{s}}  \Bigr) = J_{\lambda_{1}\ldots\lambda_{s}} .\label{LA1}
 \end{gather}
The symmetrized sums $\sum_{1}$ and $\sum_{2}$ are
over all inequivalent index permutations and have $s$ and $s(s-1)/2$
terms respectively\footnote{This is described in more detail in
Appendix~\ref{app:symmetrization}.}. 
In order to get the correct equation we have to take also the double-traceless part of
the curly bracket.
However, it will turn out that the resulting expression~(\ref{LA1}) is
already double-traceless. The fact, that the symmetrized terms in the curly bracket
in~\eqref{FronsdalVar} are already double-traceless, is a major advantage of this
Lagrange formulation. If it were not the case, we would need to project the variation to the
double-traceless part\footnote{This projection is given for any rank-$s$ symmetric tensor
f\/ield in Appendix~\ref{app:traceless}.}. 
Thus the equation of motion for the Abelian tensor gauge f\/ield $A_{\lambda_{1}
\ldots\lambda_{s}}$ is indeed the equation~(\ref{LA1}) and it contains a second order linear
dif\/ferential operator $L$ acting on the f\/ield~$A$
\begin{gather*}
(LA)_{\lambda_{1}\ldots\lambda_{s}}  \equiv  -\partial^2 A_{\lambda_{1}
\ldots\lambda_{s}}+\sum_{1}\partial_{\lambda_{1}}\partial^{\nu}A_{\nu\lambda_{2}
\ldots\lambda_{s}}-\sum_{2}\partial_{\lambda_{1}}\partial_{\lambda_{2}}A'_{\lambda_{3}
\ldots\lambda_{s}}\nonumber \\
\phantom{(LA)_{\lambda_{1}\ldots\lambda_{s}}  \equiv}{} -\sum_{2}\eta_{\lambda_{1}\lambda_{2}}\Bigl(\partial^{\mu}\partial^{\nu}
 A_{\mu\nu\lambda_{3}\ldots\lambda_{s}}-\partial^2 A'_{\lambda_{3}\ldots\lambda_{s}}-
 \tfrac{1}{2}\sum_{1}\partial_{\lambda_{3}}\partial^{\mu}A'_{\mu\lambda_{4}
 \ldots\lambda_{s}}\Bigr) , 
 \end{gather*}
whose double trace is equal to zero $(LA)^{''}_{\lambda_{5}\ldots\lambda_{s}}\equiv 0$
(see below in (\ref{FronsEomSecondTrace})).
We can express the equation (\ref{LA1}) in the operator form as
 \begin{gather}\label{FronsVarDer}
(LA)_{\lambda_{1}\ldots\lambda_{s}} =
J_{\lambda_{1}\ldots\lambda_{s}}.
\end{gather}
It follows therefore that the current also should be double traceless
\begin{gather}\label{currentdoubletarce}
J''_{\lambda_{5}\ldots\lambda_{s}} =0  .
\end{gather}
This is consistent with the observation made after formula~(\ref{hDoubleTraceless}).
These equations completely  def\/ine the theory and our intention now
is to describe the physical properties of the equation~(\ref{LA1}),~(\ref{FronsVarDer}).

Let us compute f\/irst the divergence of the l.h.s. of the
equation (\ref{LA1}) in order to check if it is equal to zero or not. This will tell
us about current divergence
$\partial^{\mu} J_{\mu\lambda_{2}\ldots\lambda_{s}}$ through the
equation of motion (\ref{LA1}), (\ref{FronsVarDer}).
The straightforward computation gives
\begin{gather}
 -\partial^{\mu}(LA)_{\mu\lambda_{2}\ldots\lambda_{s}} \nonumber\\
 \qquad  = \sum_{2}
\eta_{\lambda_{2}\lambda_{3}}\Bigl(\partial^{\mu}\partial^{\nu}
\partial^{\rho}A_{\mu\nu\rho\lambda_{4}\ldots\lambda_{s}}\!\!-
\tfrac{3}{2}\partial^{\mu} \partial^2 A'_{\mu\lambda_{4}\ldots\lambda_{s}}\!\!-
\tfrac{1}{2}\sum_{1}\partial_{\lambda_{4}}\partial^{\mu}\partial^{\nu}
A'_{\mu\nu\lambda_{4}\ldots\lambda_{s}}\Bigr),\label{FronsDiv}
\end{gather}
and it is obviously not equal to zero. Therefore the current is not conserved in a
usual sense:
$
\partial^{\mu} J_{\mu\lambda_{2}\ldots\lambda_{s}} \neq 0.
$
The full conservation gets replaced by a weaker condition which becomes
transparent after calculating the trace of the divergence:
\begin{gather}
 -\partial^{\mu} (LA)'_{\mu\lambda_{4} \ldots \lambda_{s} }
\nonumber \\
\qquad{}=(d+2s-6) \Bigl(\partial^{\mu}\partial^{\nu}\partial^{\rho}A_{\mu\nu\rho\lambda_{4}
\ldots\lambda_{s}}-\tfrac{3}{2}\partial^{\mu} \partial^2 A'_{\mu\lambda_{4}\ldots\lambda_{s}}-
\tfrac{1}{2}\sum_{1}\partial_{\lambda_{4}}\partial^{\mu}\partial^{\nu}A'_{\mu\nu\lambda_{4}
\ldots\lambda_{s}}\Bigr) .\!\!\label{FronsTrDiv}
\end{gather}
One can clearly see that
there is a simple algebraic relation between the divergence
(\ref{FronsDiv}) and trace of the divergence (\ref{FronsTrDiv})
\begin{gather}\label{tl-divergenceless}
\partial^{\mu}(LA)_{\mu\lambda_{2}\ldots\lambda_{s}}-\tfrac{1}{d+2s-6}\sum_{2}
\eta_{\lambda_{2}\lambda_{3}}\partial^{\mu}(LA)'_{\mu\lambda_{4}\ldots\lambda_{s}}=0.
\end{gather}
Because the equation of motion has the form $LA=J$, where
$J$ is the current, it follows that the equation is self-consistent and
has solutions only if the current obeys the same relation
as~$LA$, or, in other words, it has to fulf\/il a {\it weaker current conservation}
when $s \geq 3$ \cite{schwinger,fronsdal}
\begin{gather}\label{modified-current-cons}
\partial^{\mu}J_{\mu\lambda_{2}\ldots\lambda_{s}} -
\tfrac{1}{d+2s-6}\sum_{2}\eta_{\lambda_{2}\lambda_{3}}\partial^{\mu}J'_{\mu\lambda_{4}
\ldots\lambda_{s}}=0 .
\end{gather}
Thus the current is fully conserved only when $s=1,2$,
but for general $s \geq 3 $ the current is not conserved in a usual sense because only the
{\it traceless part of the current divergence vanishes}\footnote{\label{fn:dbl-tl-current}Remember that $J_{\lambda_1\ldots\lambda_s}$ is double
traceless (\ref{currentdoubletarce}).
Taking the divergence and taking the trace are commuting operations.
Therefore also $\partial^\mu J_{\mu\lambda_2\ldots\lambda_s}$ is double traceless.
The traceless part of a double traceless f\/ield $A_{\lambda_1\ldots\lambda_s}$ is given by
$
A_{\lambda_{1}\ldots\lambda_{s}}-
\tfrac{1}{d+2s-4}\sum_{2}\eta_{\lambda_{1}\lambda_{2}}
A'_{\lambda_{3}\ldots\lambda_{s}}
$.
The divergence of $J$ has only rank $s-1$ which leads to the dif\/ferent prefactor
in (\ref{modified-current-cons}).
If there were no restriction to double traceless f\/ields, the traceless part would
contain subtractions of higher traces as well. The full projection is given in Appendix~\ref{app:traceless}. 

One should
stress that the traceless part of the divergence of the current in
(\ref{modified-current-cons}) dif\/fers from the divergence of the traceless part $
\partial^{\lambda_{1}}\Bigl(J_{\lambda_{1}\ldots\lambda_{s}}-
\tfrac{1}{d+2s-4}\sum_{2}\eta_{\lambda_{1}\lambda_{2}}
J'_{\lambda_{3}\ldots\lambda_{s}}\Bigr) \neq 0
$,
which does not vanish.}.
Our main concern in the subsequent sections is
to demonstrate that this weaker current conservation law guarantees the  unitarity
of the theory. Physically this means that
only waves with
transverse polarizations are propagating very far from the sources,
as it is the case for fully conserved currents.

It is also true that the equations (\ref{tl-divergenceless}) and (\ref{modified-current-cons})
are consequences of the local gauge inva\-rian\-ce of the action~(\ref{FronsdalAction}) with
respect to the above Abelian gauge transformation of the tensor f\/ield~(\ref{generalgaugetransfromation}).
The variation of the kinetic term in the action~(\ref{FronsdalAction})
with respect to the transformation~(\ref{generalgaugetransfromation})   is
\begin{gather}\label{gaugeinvariance}
\delta_{\xi}S =
 \int d^{d}x\, (LA)_{\lambda_{1}\ldots\lambda_{s}} \delta A^{\lambda_{1}\ldots
 \lambda_{s}} =
-s\int d^{d}x\, \xi^{\lambda_{2}\ldots
\lambda_{s}}\partial^{\lambda_{1}}(LA)_{\lambda_{1}\ldots\lambda_{s}}.
 \end{gather}
If $\xi$ is traceless, then the contraction with
$\xi$ projects to the traceless part of the divergence of $LA$ which as we have
seen in~(\ref{tl-divergenceless}) vanishes, then $\delta_{\xi}S=0$.
The gauge invariance of the equation of motion  (\ref{tl-divergenceless})
and (\ref{gaugeinvariance}) and the fact that  $L$ is a linear operator implies
that any pure gauge f\/ield of the form~(\ref{generalgaugetransfromation}) is a solution of
the homogenous equation $LA =0$.  Therefore one can add to any
particular solution $A$ of~(\ref{LA1}) a pure gauge f\/ield to form a new solution
\begin{gather*}
A_{\lambda_{1}\ldots\lambda_{s}} \rightarrow  A_{\lambda_{1}\ldots\lambda_{s}} +
\sum_{1}\partial_{\lambda_{1}}\xi_{\lambda_{2}\ldots\lambda_{s}}.
\end{gather*}

Deriving the equation of motion
we have used the fact that the expression $LA$ in (\ref{FronsVarDer})
is already double traceless. To check this, notice that
the linear operator $L$   can be represented in the form
\begin{gather}\label{FronsEom}
 (LA)_{\lambda_{1}\ldots\lambda_{s}}
\equiv  (L_{0}A)_{\lambda_{1}\ldots\lambda_{s}}-\tfrac{1}{2}\sum_{2}
 \eta_{\lambda_{1}\lambda_{2}}(L_{0}A)'_{\lambda_{3}\ldots\lambda_{s}},
\end{gather}
where $L_0$ is given by
\begin{gather}\label{Lzero}
(L_{0}A)_{\lambda_{1}\ldots\lambda_{s}}   \equiv   -\partial^2
A_{\lambda_{1}\ldots\lambda_{s}}+\sum_{1}\partial_{\lambda_{1}}\partial^{\nu}
A_{\nu\lambda_{2}\ldots\lambda_{s}}-\sum_{2}\partial_{\lambda_{1}}\partial_{\lambda_{2}}
A'_{\lambda_{3}\ldots\lambda_{s}}
\end{gather}
with its trace being
\begin{gather}
(L_{0}A)'_{\lambda_{3}\ldots\lambda_{s}} = -2 \partial^2
A'_{\lambda_{3}\ldots\lambda_{s}}+2\partial^{\mu}\partial^{\nu}
A_{\mu\nu\lambda_{3}\ldots\lambda_{s}}\nonumber\\
\phantom{(L_{0}A)'_{\lambda_{3}\ldots\lambda_{s}} =}{} -\sum_{1}\partial_{\lambda_{3}}\partial^{\mu}
A'_{\mu\lambda_{4}\ldots\lambda_{s}}-\sum_{2}\partial_{\lambda_{1}}\partial_{\lambda_{2}}
A''_{\lambda_{3}\ldots\lambda_{s}}.\label{lastterm}
\end{gather}
The last term vanishes because $A^{''}=0$. Calculating the
double trace of $(L_{0}A)$, terms with single traces $A'$ of the tensor gauge f\/ield
all cancel and we get (using the fact that the last term of~(\ref{lastterm}) already vanishes)
\begin{gather*}
(L_{0}A)''_{\lambda_{5}\ldots\lambda_{s}}=-2 \partial^2
A''_{\lambda_{5}\ldots\lambda_{s}}- \sum_{1}\partial_{\lambda_{5}}\partial^{\mu}
A''_{\mu\lambda_{6}\ldots\lambda_{s}} =0  . 
\end{gather*}
Notice that
\begin{gather*}
(LA)'_{\lambda_{3}\ldots\lambda_{s}}   =   -\tfrac{1}{2}\left(d+2s-6\right)
(L_{0}A)'_{\lambda_{3}\ldots\lambda_{s}},
\end{gather*}
and therefore we have
\begin{gather}\label{FronsEomSecondTrace}
(LA)''_{\lambda_{5}\ldots\lambda_{s}}= -\tfrac{1}{2}\left(d+2s-6\right)
(L_0A)''_{\lambda_{5}\ldots\lambda_{s}}=0.
\end{gather}
In summary we have the Lagrangian (\ref{FronsdalAction}), the corresponding equations
of motion (\ref{LA1}) and a~weak current conservation (\ref{modified-current-cons}) which is
the  consequence of the invariance of the action with respect to the
Abelian gauge transformations~(\ref{generalgaugetransfromation})
with traceless gauge parameters~$\xi$.

\section[Solving the equation in de Donder-Fronsdal gauge]{Solving the equation in de Donder--Fronsdal gauge}

The idea for solving the equation of motion (\ref{LA1}) in the presence of the external  current
$J_{\lambda_{1}\ldots\lambda_{s}}$ is to f\/ind a possible gauge f\/ixing condition
imposed on the  f\/ield $A_{\lambda_{1}\ldots\lambda_{s}}$ in which the equation of motion
reduces to its diagonal form:
$-\partial^2 A_{\lambda_{1}\ldots\lambda_{s}}=(PJ)_{\lambda_{1}\ldots\lambda_{s}}$.
 In order to realize this program one should make
two important steps \cite{fronsdal}. The f\/irst step is to represent the linear dif\/ferential
operator $L$ in (\ref{FronsEom}) as a product of two  operators $R$ and $L_0$
\begin{gather*}
(R  L_{0} A)_{\lambda_{1}\ldots\lambda_{s}} = J_{\lambda_{1}\ldots\lambda_{s}},
\end{gather*}
where the operator $R$
\begin{gather}\label{B}
(R  A)_{\lambda_{1}\ldots\lambda_{s}}   \equiv
A_{\lambda_{1}\ldots\lambda_{s}}-\tfrac{1}{2}\sum_{2}\eta_{\lambda_{1}\lambda_{2}}
A'_{\lambda_{3}\ldots\lambda_{s}}
\end{gather}
is a nonsingular algebraic operator with its inverse $P$
\begin{gather}\label{A}
(P A)_{\lambda_{1}\ldots\lambda_{s}}   \equiv
  A_{\lambda_{1}\ldots\lambda_{s}}-\tfrac{1}{(d+2s-6)}\sum_{2}\eta_{\lambda_{1}\lambda_{2}}
 A'_{\lambda_{3}\ldots\lambda_{s}}
\end{gather}
 and $L_{0}$ is the second order dif\/ferential operator given in~(\ref{Lzero}).
The second step is to represent the operator $L_0$ in the following form
\begin{gather*}
(L_{0}  A)_{\lambda_{1}\ldots\lambda_{s}}=
-\partial^2 A_{\lambda_{1}\ldots\lambda_{s}}+\sum_{1}\partial_{\lambda_{1}}
\Bigl(\partial^{\nu}A_{\nu\lambda_{2}\ldots\lambda_{s}}-
\tfrac{1}{2}\sum_{1}\partial_{\lambda_{2}}
A'_{\lambda_{3}\ldots\lambda_{s}}\Bigr).
\end{gather*}
From the last expression one can deduce that if we could impose the gauge
condition on the gauge f\/ield $A$ of the form
\begin{gather}\label{genDeDonderI}
\partial^{\mu}A_{\mu\lambda_{2}\ldots\lambda_{s}}-
\tfrac{1}{2}\sum_{1}\partial_{\lambda_{2}}A'_{\lambda_{3}\ldots\lambda_{s}}=0
\end{gather}
then the operator $L_{0}$ would reduce to the d'Alembertian:
\begin{gather*}
(L_{0} A)_{\lambda_{1}\ldots\lambda_{s}}=-\partial^2
A_{\lambda_{1}\ldots\lambda_{s}},
\end{gather*}
and the equation  of motion $R L_{0} A = R (-\partial^2)A =J$   can be solved by using the inverse operator~$P$. Thus we have
\begin{gather*}
-\partial^2 A_{\lambda_{1}\ldots\lambda_{s}}   =
(PJ)_{\lambda_{1}\ldots\lambda_{s}}.
\end{gather*}
In momentum space, where $-\partial^2\To k^{2}$, a solution to the
above equation is given by the formula
\begin{gather}\label{Frons:expl-h-solution}
A_{\lambda_{1}\ldots\lambda_{s}}= \frac{(PJ)_{\lambda_{1}\ldots\lambda_{s}} } {k^{2}}.
\end{gather}

The crucial question about the gauge f\/ixing condition (\ref{genDeDonderI})  is,   whether
it is accessible or not. Let us see  how that expression transforms under the gauge transformation
(\ref{genDeDonderI})
\begin{gather*}
\delta_{\xi}\Bigl(\partial^{\mu}A_{\mu\lambda_{2}\ldots\lambda_{s}}-
\tfrac{1}{2}\sum_{1}\partial_{\lambda_{2}}A'_{\lambda_{3}\ldots\lambda_{s}}\Bigr)  =
\square\xi_{\lambda_{2}\ldots\lambda_{s}}.
\end{gather*}
It is obvious that, if the l.h.s.\ is not equal to zero,
then one can always f\/ind a solution $\xi$ so as to fulf\/il the gauge condition~(\ref{genDeDonderI}).
Let us call it de Donder--Fronsdal  gauge, because for $s=2$
it coincides with  de Donder gauge in gravity\footnote{In contrast to the gravity case $s=2$,
the gauge f\/ixing condition (\ref{genDeDonderI}) for general $s$ cannot be
written as the divergence of
$A_{\lambda_{1}\ldots\lambda_{s}}-
\tfrac{1}{2}\sum_{2}\eta_{\lambda_{1}\lambda_{2}}A{}'_{\lambda_{3}\ldots\lambda_{s}}$.
However, it can be written as the traceless part of its divergence
\[
\proj_{\rho_{2}\ldots\rho_{s}}\hoch{\lambda_{2}\ldots
\lambda_{s}}\partial^{\lambda_{1}}\Bigl(A_{\lambda_{1}\ldots\lambda_{s}}-
\tfrac{1}{2}\sum_{2}\eta_{\lambda_{1}\lambda_{2}}A{}'_{\lambda_{3}
\ldots\lambda_{s}}\Bigr)=0 .\nn
\]
Here $\Pi$ is the projector to the traceless part given in Appendix~\ref{app:traceless}.}. 

\section[Interaction of higher spin field with external currents]{Interaction of higher spin f\/ield with external currents}

With the solution (\ref{Frons:expl-h-solution}) at hand we can f\/ind out the properties
of the f\/ield $A$ propagating far from the current $J$ when the latter is constrained to be
weakly conserved (\ref{modified-current-cons}).
The main result of \cite{schwinger,fronsdal} is
that only transverse degrees of freedom propagate to inf\/inity, even when
the current is only weakly conserved (\ref{modified-current-cons}).
For completeness let us recollect the corresponding results for the
lower-rank gauge f\/ields
\cite{feynman,schwinger} and then present the proof of \cite{fronsdal}
for the general case.

In electrodynamics ($s=1$) and linearized gravity ($s=2$)  (\ref{electromaggaravity}) the currents
are fully conserved
\begin{gather}\label{fullcurrentconservatio}
k^{\mu}J_{\mu}=0,\qquad k^{\mu}J_{\mu\nu}=0,
\end{gather}
and the interaction between currents can be straightforwardly analyzed  \cite{feynman}.
But already for the Schwinger equation of rank-3 gauge f\/ields  the weaker
conservation (\ref{modified-current-cons}) takes place~\cite{schwinger}
\begin{gather}\label{weakcurrentconservation}
k^{\mu}J_{\mu\nu\lambda}-{1\over d} \eta_{\nu\lambda} k^{\mu} J^{'}_{\mu} = 0.
\end{gather}
Thus we have to consider two cases: when the currents are fully conserved
(\ref{fullcurrentconservatio}) and the case when it is weakly conserved~(\ref{weakcurrentconservation}).

In the general action (\ref{FronsdalAction}) the interaction term of the gauge f\/ield with the current
$\tilde{J}$ is of the form $-A \tilde{J} $, therefore the exchange interaction between
two currents $J$ and $\tilde{J}$ can
be found with the help of the gauge f\/ield generated by a source $J$ in~(\ref{Frons:expl-h-solution})
\begin{gather}\label{interaction}
-  A_{\rho_1 \ldots  \rho_s} \tilde{J}^{\rho_1 \ldots  \rho_s} = -
 J^{\lambda_1 \ldots  \lambda_s} {P_{\lambda_1 \ldots  \lambda_s , \rho_1 \ldots  \rho_s}\over k^2}
\tilde{J}^{\rho_1 \ldots  \rho_s}
\end{gather}
with the expression
\begin{gather*}
\Delta_{\lambda_1 \ldots  \lambda_s , \rho_1 \ldots  \rho_s}(k) =
{P_{\lambda_1 \ldots  \lambda_s , \rho_1 \ldots  \rho_s}\over k^2}
\end{gather*}
representing the propagator of the rank-$s$ gauge f\/ield. The symmetric operator $P$ is
given by~(\ref{A}). For the lower-rank f\/ields the interaction has the following form
\begin{gather}
  s=1\quad -J^{\lambda}\frac{\eta_{\lambda\rho}}{k^{2}}\tilde{J}^{\rho}, \label{MW-prop}\\
  s=2\quad -J^{\lambda_{1}\lambda_{2}} \frac{ \eta_{\lambda_{1}\rho_{1}}\eta_{\lambda_{2}\rho_{2}}-
\tfrac{1}{d-2}\eta_{\lambda_{1}\lambda_{2}}\eta_{\rho_{1}\rho_{2}}
}{k^{2}}  \tilde{J}^{\rho_{1}\rho_{2}},  \label{grav-prop}\\
  s=3\quad - J^{\lambda_{1}\lambda_{2}\lambda_{3}} \frac{
   \eta_{\lambda_{1}\rho_{1}}\eta_{\lambda_{2}\rho_{2}}\eta_{\lambda_{3}\rho_{3}}-
\tfrac{3}{d}  \eta_{\lambda_{1}\rho_{1}} \eta_{\lambda_{2}\lambda_{3}} \eta_{\rho_{2} \rho_{3}}
}{k^{2}}   \tilde{J}^{\rho_{1}\rho_{2}\rho_{3}}. \label{schwinger}
\end{gather}
To simplify the analysis of this interaction we can always take
the momentum vector $k$ in the form:
\begin{gather*}
k_{\mu}=(\omega,0,\ldots,0, \kappa)
\end{gather*}
and introduce the parity reversed momentum vector
\begin{gather*}
\bar{k}_{\mu}=(\omega,0,\ldots,0,-\kappa)
\end{gather*}
together with $d-2$ space-like orthogonal vectors $e^{\mu}_{i}$, $i=1,\dots,d-2$:
\begin{gather*}
e^{\mu}_{1} =  (0,1,...0,0),\\
\cdots\cdots\cdots\cdots\cdots\cdots\\
e^{\mu}_{d-2} =  (0,0,\dots,1,0).
\end{gather*}
These vectors form a frame and the metric tensor can be represented in the form
\begin{gather}\label{c}
\eta^{\mu\nu}   =   -\sum_{i}e^{\mu}_{i} e^{\nu}_{i}   +
\frac{(k + \bar{k})^{\mu}(k + \bar{k})^{\nu}}{2(k^2 + k\bar{k})}+
\frac{(k - \bar{k})^{\mu}(k - \bar{k})^{\nu}}{2(k^2 - k\bar{k})},
\end{gather}
where the f\/irst term projects  to the transversal plane, while
the remaining ones project to the longitudinal direction. On the
mass-shell $k^2 = \bar{k}^2 =\omega^2 -\kappa^2 =0$ this expression
reduces to the familiar expression \cite{schwinger}
\begin{gather}\label{onmassshell}
\eta^{\mu\nu}   =  - \sum_{i}e^{\mu}_{i} e^{\nu}_{i}   +
\frac{k^{\mu}\bar{k}^{\nu}  + \bar{k}^{\mu}k^{\nu}}{k\bar{k} }.
\end{gather}
Armed with the last two expressions one can prove that only transversal
polarizations of the tensor gauge boson participate in the exchange interaction
between currents at large distances, when
$k^2 \approx 0$. Indeed, inserting the representation~(\ref{onmassshell})
into~(\ref{MW-prop}) and~(\ref{grav-prop}) and using the current conservation,
which is valid in these cases~(\ref{fullcurrentconservatio}), we shall get
\begin{gather*}
  s=1\quad -J^{\lambda}\frac{\eta_{\lambda\rho}}{k^{2}}\tilde{J}^{\rho} =
  \frac{ J_{\lambda}e^{\lambda}_i  e^{\rho}_i \tilde{J}_{\rho} }{\omega^2 -\kappa^2} =
  \frac{ J_{i} \tilde{J}_{i} }{\omega^2 -\kappa^2},\\
  s=2\quad -  J^{\lambda_{1}\lambda_{2}}\frac{ \eta_{\lambda_{1}\rho_{1}}\eta_{\lambda_{2}\rho_{2}}-
\tfrac{1}{d-2}\eta_{\lambda_{1}\lambda_{2}}\eta_{\rho_{1}\rho_{2}}
}{k^{2}} \tilde{J}^{\rho_{1}\rho_{2}} \\
\phantom{s=2\quad}{} =  - \frac{ J_{\lambda_{1}\lambda_{2}} e^{\lambda_{1}}_{i}  e^{\lambda_{2}}_{j} ~ ~
e^{\rho_{1}}_{i}  e^{\rho_{2}}_{j}  \tilde{J}_{\rho_{1}\rho_{2}}-
\tfrac{1}{d-2} J_{\lambda_{1}\lambda_{2}} e^{\lambda_{1}}_{i}e^{\lambda_{2}}_{i}
e^{\rho_{1}}_{j}e^{\rho_{2}}_{j} \tilde{J}_{\rho_{1}\rho_{2}} }{\omega^{2}-\kappa^{2}}  =  -\frac{J_{ij} \tilde{J}_{ij}-
 \tfrac{1}{d-2} J_{ii} \tilde{J}_{jj}} {\omega^{2}-\kappa^{2}}.
\end{gather*}
All bilinear terms $k_{\mu}\bar{k}_{\nu}$   are cancelled
because  of the  current conservation,  and quantities
\begin{gather*}
J_{i}=J_{\lambda } e_i^{\lambda },\qquad J_{ij}=J_{\lambda_{1}\lambda_{2}} e^{\lambda_{1}}_{i}e^{\lambda_{2}}_{j}
\end{gather*}
are projection of currents to the transverse plane.
At the pole $\omega^{2}-\kappa^2 =0$ the residues
are positive def\/inite. Indeed, for $s=1$ we have $J_i J_i$ and for $s=2$ the numerator can be written
as a square of the traceless part of $J_{ij}$
\begin{gather*}
 \left( J_{ij}-\tfrac{1}{d-2}\delta_{ij} J_{nn} \right)
 \left( J_{ij}-\tfrac{1}{d-2}\delta_{ij} J_{mm} \right).
\end{gather*}
It is obvious, how to extended this proof to the higher-rank f\/ields, if the corresponding currents would be
fully conserved, but unfortunately they are not!  What is amazing nevertheless, is that for
weakly conserved currents (\ref{modified-current-cons}), (\ref{weakcurrentconservation})
the analysis can be reduced to the case of fully conserved currents.
Therefore it is worth to follow the general Schwinger consideration of the exchange interaction between
conserved currents~\cite{schwinger}.
The general form of the exchange interaction (\ref{interaction}) is
\begin{gather}
 -  J^{\lambda_1 \ldots  \lambda_s} {P_{\lambda_1 \ldots  \lambda_s , \rho_1 \ldots  \rho_s}\over k^2}
\tilde{J}^{\rho_1 \ldots  \rho_s}\nonumber\\
\qquad{} =-J^{\lambda_{1}\ldots\lambda_{s}}\frac{
 \left(\eta_{\lambda_{1}\rho_{1}}\ldots\eta_{\lambda_{s}\rho_{s}}-
 \tfrac{s(s-1)}{2(d+2s-6)}\eta_{\lambda_{1}\lambda_{2}}\eta_{\rho_{1}\rho_{2}}
 \eta_{\lambda_{3}\rho_{3}}\ldots\eta_{\lambda_{s}\rho_{s}}\right)
 }{k^{2}}  \tilde{J}^{\rho_{1}\ldots\rho_{s}}  ,\label{exchangeinteraction}
\end{gather}
where we have used the expression for the matrix $P$ in~(\ref{A}).
Again inserting the representation (\ref{onmassshell})
for the metric tensor into the (\ref{exchangeinteraction})  and supposing
that the currents are conserved:
$k^{\mu}J_{\mu\lambda_{2}\ldots\lambda_{s}}=0,~k^{\mu}\tilde{J}_{\mu\lambda_{2}\ldots\lambda_{s}}=0$,
we shall get
\begin{gather*}
(-)^{s+1} \frac{J_{i_{1}\ldots i_{s}}  \tilde{J}_{i_{1}\ldots i_{s}}
- \tfrac{s(s-1)}{2(d+2s-6)}   J^{'}_{i_{3}\ldots i_{s}}  \tilde{J}^{'}_{i_{3}\ldots i_{s}}
 }{\omega^2 - \kappa^{2}},
 \end{gather*}
 where
\begin{gather*}
J_{i_{1}\ldots i_{s}} = J_{\lambda_1 \ldots  \lambda_s} e^{\lambda_1}_{i_1} \cdots e^{\lambda_s}_{i_s}.
\end{gather*}
The longitudinal modes $k_{\mu}\bar{k}_{\nu}$ do not contribute because of the
current conservation and we are left with only transversal propagating modes!
The expression in the above equation
coincides  with the product of the traceless parts of the currents, as it was  for
$s=2$. 
Indeed, the trace  has reduced to the transversal directions,
and the ef\/fective dimension has therefore reduced by~2, and the coef\/f\/icient $\tfrac{s(s-1)}{2(d+2s-6)}$
is the correct coef\/f\/icient for the traceless projector for f\/ields of rank $s$
in dimension $d-2$.\footnote{Compare with Appendix~\ref{app:traceless}.}
Our main concern in the next section is to prove that almost the same mechanism works in the case
of the weakly conserved currents~\cite{schwinger,fronsdal}.

In the above discussion we have considered interactions at large distances, when
$k^2 = \omega^2 - \kappa^2 \approx 0$, therefore keeping the most singular terms.
In order to analyze  the short distance behaviour, when $\omega^{2}-k^2 \neq 0$,
one should use the relation~(\ref{c}) and follow the beautiful consideration of
Feynman~\cite{feynman}.

\section{Interaction of weakly conserved currents}
\label{sec:interaction-weakly}

In order to prove that in the case of weakly conserved currents the propagating
modes are  positive def\/inite transversal polarizations we have
to reformulate the exchange interaction (\ref{interaction})
\begin{gather*}
-J^{\lambda_1 \ldots  \lambda_s} {P_{\lambda_1 \ldots  \lambda_s , \rho_1 \ldots  \rho_s}\over k^2}
\tilde{J}^{\rho_1 \ldots  \rho_s}
\end{gather*}
in a way that it becomes \cite{fronsdal}
\begin{gather*}
-J_{ f}^{~~\lambda_1 \ldots  \lambda_s} {P_{\lambda_1 \ldots  \lambda_s , \rho_1 \ldots  \rho_s}\over k^2}
\tilde{J}^{~~\rho_1 \ldots  \rho_s}_{ f},
\end{gather*}
where the ef\/fective current $J_{ f}$ is fully conserved.
Let us introduce the projection $\proj$ to the traceless part, which we already used implicitly
several times. Its action on double traceless tensor $t_{\lambda_1\ldots\lambda_{s-1}}$ of rank $s-1$
is given by\footnote{As described in Appendix~\ref{app:traceless}, higher traces appear in
the projection, if $A$ is not double traceless.}
\begin{gather*}
\proj_{\lambda_1\ldots\lambda_{s-1}}\hoch{\rho_1\ldots\rho_{s-1}}
t_{\rho_1\ldots\rho_{s-1}}=t_{\rho_1\ldots\rho_{s-1}}-
\tfrac{1}{d+2s-6}\sum_2\eta_{\rho_1\rho_2}t'_{\rho_3\ldots\rho_{s-1}},
\end{gather*}
and we can represent the weak current conservation (\ref{modified-current-cons})
in the following  form
\begin{gather*}
\Pi^{\lambda_{2}\ldots\lambda_{s}}\tief{\rho_{2}\ldots\rho_{s}} k_{\mu}J^{\mu\rho_{2}
\ldots\rho_{s}}  =   0,
\end{gather*}
with $t_{\lambda_2\ldots\lambda_{s}}= k_{\mu}J^{\mu\rho_{2} \ldots\rho_{s}}$. This equation
can be contracted  with an arbitrary tensor $f_{\lambda_{2}\ldots\lambda_{s}}$ of rank $s-1$,
and because $\Pi$ is a symmetric matrix  this can be written as
\begin{gather}\label{cancellation}
k_{\rho_{1}} (\Pi f)_{\rho_{2}\ldots\rho_{s}} J^{\rho_{1}\ldots\rho_{s}}  =  0 ,
\end{gather}
that is, the contraction of ${1\over s}\sum_1k_{\rho_{1}} (\Pi f)_{\rho_{2}\ldots\rho_{s}}$ with the current
vanishes for all~$f$. The interpretation of this formula is, that instead of the longitudinal
operator $k_{\rho_{1}}$ in the case of fully conserved currents, we have
the operator ${1\over s}\sum_1k_{\rho_{1}} (\Pi f)_{\rho_{2}\ldots\rho_{s}}$
which plays a similar role.

Now one can add this operator to the current $J$ to form an ef\/fective current $J_{f}$
\begin{gather}\label{effectivecurrent}
J_{f}^{~~\lambda_1 \ldots  \lambda_s} = J^{\lambda_1 \ldots  \lambda_s}+
R^{\lambda_{1}\ldots\lambda_{s},\rho_{1}\ldots\rho_{s}}
k_{\rho_{1}} (\Pi f)_{\rho_{2}\ldots\rho_{s}},
\end{gather}
where $R$ was def\/ined in~(\ref{B}).
The interaction of the ef\/fective currents $J_{ f} P \tilde{J}_{ f}$ will
be identical with the original interaction of currents $J  P \tilde{J} $, if the cross terms and the
square of the additional operator vanish. The cross terms will vanish, because they simply
express the weak current conservation~(\ref{cancellation}). For the square we have
\begin{gather*}
k^{\rho_{1}} (\Pi f)^{\rho_{2}\ldots\rho_{s}}
R_{\rho_{1}\ldots\rho_{s},\lambda_{1}\ldots\lambda_{s}}
k^{\lambda_{1}} (\Pi \tilde f)^{\lambda_{2}\ldots\lambda_{s}} \\
=\tfrac{1}{s}\sum_1 k^{\rho_{1}} (\Pi f)^{\rho_{2}\ldots\rho_{s}}
\left(\eta_{\rho_{1}\lambda_{1}}\cdots\eta_{\rho_{s}\lambda_{s}}-
 \tfrac{s(s-1)}{4}\eta_{\rho_{1}\rho_{2}}\eta_{\lambda_{1}\lambda_{2}}
 \eta_{\rho_{3}\lambda_{3}}\cdots\eta_{\rho_{s}\lambda_{s}}\right)
\tfrac{1}{s}\sum_1 k^{\lambda_{1}} (\Pi \tilde f)^{\lambda_{2}\ldots\lambda_{s}}\\
=\tfrac{1}{s} k^{2} (\Pi f)_{\lambda_1\ldots\lambda_s} (\Pi \tilde{f})^{\lambda_1\ldots\lambda_s},
\end{gather*}
where we have used the fact that $(\Pi f)$ is traceless. It vanishes on the mass-shell
$k^{2}=0$. Therefore we have
\begin{gather*}
-J_{f}^{~~\lambda_1 \ldots  \lambda_s}~
{P_{\lambda_1 \ldots  \lambda_s , \rho_1 \ldots  \rho_s}\over k^2}
\tilde{J}^{~~\rho_1 \ldots  \rho_s}_{f}
=-(J +  k(\Pi f) R~)^{\lambda_1 \ldots  \lambda_s}
{P_{\lambda_1 \ldots  \lambda_s , \rho_1 \ldots  \rho_s}\over k^2}
(\tilde{J} +  R  k(\Pi \tilde{f}) )^{\rho_1 \ldots  \rho_s}
\\
\qquad{}=-J^{\lambda_1 \ldots  \lambda_s} {P_{\lambda_1 \ldots  \lambda_s , \rho_1 \ldots  \rho_s}\over k^2}
\tilde{J}^{\rho_1 \ldots  \rho_s}
-{k_{\rho_1} (\Pi f)_{\rho_2 \ldots  \rho_s} \tilde{J}^{\rho_1 \ldots  \rho_s} \over k^2}
-{ J_{\rho_1 \ldots  \rho_s}
k^{\rho_1}(\Pi \tilde{f})^{\rho_2 \ldots  \rho_s} \over k^2}\\
\qquad{}- k^{\rho_{1}} (\Pi f)^{\rho_{2}\ldots\rho_{s}}
{R_{\rho_{1}\ldots\rho_{s},\lambda_{1}\ldots\lambda_{s}} \over k^2 }
k^{\lambda_{1}} (\Pi \tilde{f})^{\lambda_{2}\ldots\lambda_{s}} .
 \end{gather*}
The last three terms are equal to zero, as we already explained, and the equivalence of the
interaction  has been demonstrated with the ef\/fective current (\ref{effectivecurrent}).
Let us calculate now the divergence of the ef\/fective current
     \begin{gather*}
 k_{\lambda_1} J_{f}^{~~\lambda_1 \ldots  \lambda_s}
 =  k_{\lambda_1} J^{\lambda_1 \ldots  \lambda_s}
 +k_{\lambda_1} R^{\lambda_{1}\ldots\lambda_{s},\rho_{1}\ldots\rho_{s}}
k_{\rho_{1}} (\Pi f)_{\rho_{2}\ldots\rho_{s}} \\
\phantom{k_{\lambda_1} J_{f}^{~~\lambda_1 \ldots  \lambda_s}}{} =
\tfrac{1}{d+2s-6}\sum_{2}\eta^{\lambda_{2}\lambda_{3}}k_{\mu}J^{'\mu\lambda_{4}
\ldots\lambda_{s}} + \tfrac{1}{s} k^2 (\Pi f)^{\lambda_{2}\ldots\lambda_{s}}-
\tfrac{1}{s}\sum_2 \eta^{\lambda_2\lambda_3} k_{\mu}k_{\nu} (\Pi f)^{\mu\nu\lambda_4\dots \lambda_s}
\\
\phantom{k_{\lambda_1} J_{f}^{~~\lambda_1 \ldots  \lambda_s}}{}=\tfrac{1}{d+2s-6}\sum_{2}\eta^{\lambda_{2}\lambda_{3}}k_{\mu}J^{'\mu\lambda_{4}
\ldots\lambda_{s}} -
\tfrac{1}{s}\sum_2 \eta^{\lambda_2\lambda_3} k_{\mu}k_{\nu} (\Pi f)^{\mu\nu\lambda_4\dots \lambda_s}.
  \end{gather*}
Choosing a tensor $f_{\lambda_{2}\ldots\lambda_{s}}$
so that\footnote{We will provide an explicit solution of this equation in Appendix~\ref{app:access}.}
\begin{gather}
\tfrac{1}{s} k_{\nu} (\Pi f)^{\nu\mu\lambda_4 \dots \lambda_s}= \tfrac{1}{(d+2s-6)} J^{'\mu\lambda_{4}
\ldots\lambda_{s}},\label{tobeshown}
\end{gather}
we can get a conserved (on mass-shell) ef\/fective current
\begin{gather*}
k_{\lambda_1} J_{ f}^{~~\lambda_1 \ldots  \lambda_s}=0.
\end{gather*}
Thus the interaction $JP\tilde{J}$ can be reduced to the form $J_{ f}P\tilde{J}_{ f}$,
where $J_{ f}$ is a conserved current and the problem reduces to
the one that we already solved in the previous
section\footnote{A discussion of current interaction in the unconstrained
formulation can be found in~\cite{Francia:2007qt}.}.

Summarizing we have to stress that the self-consistency of the
equations of motion~(\ref{LA1}),~(\ref{FronsVarDer}) of this theory requires the
existence of a double-traceless~(\ref{currentdoubletarce}) and
\emph{weakly conserved} current~(\ref{modified-current-cons}).
It is outside of the  scope of this theory
to answer the question whether such external currents exist. It
should be answered by a fully interacting theory, a subject of current
research in higher spin f\/ield theory.
In this context we have to remark also that the above consideration does
not contradict the Weinberg argument on the non-existence
of a \emph{fully conserved} higher rank current of a specif\/ic form
\cite[p.~538]{Weinberg:1995mt}, \cite{Weinberg:1964cn}.

\section[Alternative representation of Schwinger-Fronsdal action]{Alternative representation of Schwinger--Fronsdal action}

The action (\ref{FronsdalAction}) is a generalization of ordinary Abelian
gauge theory and it is therefore tempting to try to write it as a square
of some f\/ield strength tensor. Field strength tensors are characterized by the property
that they transform homogenously under gauge transformations.
This means for Abelian gauge transformations that they should not transform at all.
Indeed such objects can be constructed, but unfortunately they need to be of $s$-th
derivative order~\cite{de Wit:1979pe}. This is like in gravity, where the curvature
is of second derivative order of the metric. Square of such objects for $s>1$ certainly
cannot coincide with the second order action~(\ref{FronsdalAction}).
Nevertheless they can be used to construct a nonlocal geometric action
which is related to Fronsdal's upon partial gauge f\/ixing~\cite{Francia:2002aa,Francia:2002pt}.

We shall try to write Schwinger--Fronsdal's
Lagrangian as a square, or at least as a sum of squares, of objects
that reduce for $s=1$ to the ordinary f\/ield strength tensor. To this end,
let us def\/ine \begin{gather*}
F_{\mu\nu,\lambda_{2}\ldots\lambda_{s}}  \equiv  \partial_{\mu}
A_{\nu\lambda_{2}\ldots\lambda_{s}}-\partial_{\nu}A_{\mu\lambda_{2}\ldots\lambda_{s}},\\
F'_{\mu\nu,\lambda_{4}\ldots\lambda_{s}}  \equiv  \partial_{\mu}
A'_{\nu\lambda_{4}\ldots\lambda_{s}}-\partial_{\nu}A'_{\mu\lambda_{4}\ldots\lambda_{s}},\\
H_{\mu,\lambda_{3}\ldots\lambda_{s}}  \equiv  \partial^{\nu}
A_{\nu\mu\lambda_{3}\ldots\lambda_{s}}-\tfrac{s}{2}\partial_{\mu}
A'_{\lambda_{3}\ldots\lambda_{s}}.
\end{gather*}
Their squares read
\begin{gather*}
F_{\mu\nu,\lambda_{2}\ldots\lambda_{s}}F^{\mu\nu,\lambda_{2}\ldots\lambda_{s}}  =
\left(\partial_{\mu}A_{\nu\lambda_{2}\ldots\lambda_{s}}-\partial_{\nu}A_{\mu\lambda_{2}
\ldots\lambda_{s}}\right)\left(\partial^{\mu}A^{\nu\lambda_{2}\ldots\lambda_{s}}-
\partial^{\nu}A^{\mu\lambda_{2}\ldots\lambda_{s}}\right) \\
\phantom{F_{\mu\nu,\lambda_{2}\ldots\lambda_{s}}F^{\mu\nu,\lambda_{2}\ldots\lambda_{s}}}{}
=  2\partial_{\mu}A_{\nu\lambda_{2}\ldots\lambda_{s}}\partial^{\mu}A^{\nu\lambda_{2}
 \ldots\lambda_{s}}-2\partial_{\mu}A_{\nu\lambda_{2}\ldots\lambda_{s}}
\partial^{\nu}A^{\mu\lambda_{2}\ldots\lambda_{s}}, \\
F'_{\mu\nu,\lambda_{4}\ldots\lambda_{s}}F'^{\mu\nu,\lambda_{4}\ldots\lambda_{s}}  =
\left(\partial_{\mu}A'_{\nu\lambda_{4}\ldots\lambda_{s}}-\partial_{\nu}
A'_{\mu\lambda_{4}\ldots\lambda_{s}}\right)\left(\partial^{\mu}
A'^{\nu\lambda_{4}\ldots\lambda_{s}}-\partial^{\nu}A'^{\mu\lambda_{4}\ldots\lambda_{s}}\right) \\
\phantom{F'_{\mu\nu,\lambda_{4}\ldots\lambda_{s}}F'^{\mu\nu,\lambda_{4}\ldots\lambda_{s}}}{} =  2\partial_{\mu}A'_{\nu\lambda_{4}\ldots\lambda_{s}}\partial^{\mu}
 A'^{\nu\lambda_{4}\ldots\lambda_{s}}-2\partial_{\mu}
 A'_{\nu\lambda_{4}\ldots\lambda_{s}}\partial^{\nu}A'^{\mu\lambda_{4}\ldots\lambda_{s}},  \\
H_{\mu,\lambda_{3}\ldots\lambda_{s}}H^{\mu,\lambda_{3}\ldots\lambda_{s}}  =
\left(\partial^{\nu}
A_{\nu\mu\lambda_{3}\ldots\lambda_{s}}-\tfrac{s}{2}\partial_{\mu}
A'_{\lambda_{3}\ldots\lambda_{s}}\right) \left(\partial_{\rho}
A^{\rho\mu\lambda_{3}\ldots\lambda_{s}}-\tfrac{s}{2}\partial^{\mu}
A'^{\lambda_{3}\ldots\lambda_{s}}\right) \\
\phantom{H_{\mu,\lambda_{3}\ldots\lambda_{s}}H^{\mu,\lambda_{3}\ldots\lambda_{s}} }{}  =  \partial^{\nu}A_{\nu\mu\lambda_{3}\ldots\lambda_{s}}\partial_{\rho}
 A^{\rho\mu\lambda_{3}\ldots\lambda_{s}}-s\partial^{\nu}
 A_{\nu\mu\lambda_{3}\ldots\lambda_{s}}\partial^{\mu}A'^{\lambda_{3}\ldots\lambda_{s}}\\
\phantom{H_{\mu,\lambda_{3}\ldots\lambda_{s}}H^{\mu,\lambda_{3}\ldots\lambda_{s}}  = }{} +\left(\tfrac{s}{2}\right)^{2}\partial_{\mu}
 A'_{\lambda_{3}\ldots\lambda_{s}}\partial^{\mu}A'^{\lambda_{3}\ldots\lambda_{s}},
 \end{gather*}
and the Schwinger--Fronsdal action (\ref{FronsdalAction}) can therefore be written
as
\begin{gather*}
S=\int dx^{d}\, \tfrac{1}{4}F^{2}-\tfrac{s-1}{2}H^{2}+\tfrac{s(s-1)(s-2)}{16}F'^{2}.
\end{gather*}
Despite the fact that these f\/ield strength tensors do not transform homogeneously,
the sum does. Similar f\/ield strength tensors have been introduced in~\cite{fierz} and
recently in~\cite{Savvidy:2005fi,Savvidy:2005zm,Savvidy:2005ki,Barrett:2007nn}.

\appendix
\section[Accessibility of a conserved effective current]{Accessibility of a conserved ef\/fective current}
\label{app:access}

An essential ingredient of the construction of a conserved ef\/fective current in Section~\ref{sec:interaction-weakly} was the claim in (\ref{tobeshown}) that one can choose a tensor $f$ such that
\begin{gather}
\label{claim}
k_{\mu} (\Pi f)^{\mu\lambda_3\dots \lambda_s}= \frac{s}{(d+2s-6)} J'^{\lambda_{3}\ldots\lambda_{s}} .
\end{gather}
The solution can be derived by expanding the tensors $f$ and $J'$ in the basis $e_i^\mu$,
$k^\mu$ and $\bar k^\mu$ and then compare the coef\/f\/icients on both sides.
For $s=3$ where~$f$ has rank~2 and $J'$ has rank 1 this yields the following result
\begin{gather}
\tfrac{d}{3}f^{\mu\nu}=\frac{2\bar k_\rho J'^{\rho}}{(k\bar k)^2}
\left(k^\mu\bar k^\nu+\bar k^\mu k^\nu\right)+\frac{k_\rho J'^{\rho}}{(k\bar k)^2}
\bar k^\mu \bar k^\nu+\sum_i\frac{J'_i}{k\bar k}\left(\bar k^\mu e_i^\nu+e_i^\mu
\bar k^\nu\right)-\eta^{\mu\nu}\frac{\bar k_\rho J'^\rho}{k\bar k}.\label{fresultforsisthree}
\end{gather}
In order to avoid too many prefactors in the following, let us introduce
the symbol $X$ for the tensor on the righthand side of~(\ref{claim})
\begin{gather*}
\tfrac{s}{d+2s-6}J'^{\lambda_{1}\ldots\lambda_{s-2}}\equiv X^{\lambda_{1}\ldots\lambda_{s-2}}.
\end{gather*}
For general $s$, the expansion of $f$ and $X$ in the basis takes
the form \begin{gather}
f_{\lambda_{1}\ldots\lambda_{s-1}}  =  \sum_{n_{k}=0}^{s-1}\sum_{n_{\bar{k}}=0}^{s-1-n_{k}}f_{(n_{k})
(n_{\bar{k}})i_{1}\ldots i_{s-1-n_{k}-n_{\bar{k}}}}\nonumber \\
\phantom{f_{\lambda_{1}\ldots\lambda_{s-1}}  =}{}  \times\sum_{(n_{k})}\big(k_{\lambda_{1}}\cdots k_{\lambda_{n_{k}}}\big)\sum_{(n_{\bar{k}})}
 \big(\bar{k}_{\lambda_{n_{k}+1}}\cdots\bar{k}_{\lambda_{n_{k}+n_{\bar{k}}}}\big)
 e_{\lambda_{n_{k}+n_{\bar{k}}+1}}^{i_{1}}\cdots e_{\lambda_{s-1}}^{i_{s-1-n_{k}-n_{\bar{k}}}},\label{f-exp}\\
X_{\lambda_{1}\ldots\lambda_{s-2}}  =  \sum_{n_{k}=0}^{s-2}\sum_{n_{\bar{k}}=0}^{s-2-n_{k}}X_{(n_{k})
(n_{\bar{k}})i_{1}\ldots i_{s-2-n_{k}-n_{\bar{k}}}}\nonumber \\
\phantom{X_{\lambda_{1}\ldots\lambda_{s-2}}  =}{}  \times\sum_{(n_{k})}\big(k_{\lambda_{1}}\cdots k_{\lambda_{n_{k}}}\big)\sum_{(n_{\bar{k}})}
 \big(\bar{k}_{\lambda_{n_{k}+1}}\cdots\bar{k}_{\lambda_{n_{k}+n_{\bar{k}}}}\big)
 e_{\lambda_{n_{k}+n_{\bar{k}}+1}}^{i_{1}}\cdots e_{\lambda_{s-2}}^{i_{s-2-n_{k}-n_{\bar{k}}}}.
 \label{X-exp}
 \end{gather}
The symmetrized sums $\sum_{(n)}$ are over all inequivalent index
permutations and are discussed in Appendix~\ref{app:symmetrization}. Let us f\/irst f\/ind the general
solution for the equation for unrestricted $f$
\begin{gather}
k_{\mu}f^{\mu\lambda_{2}\ldots\lambda_{s-1}}=X^{\lambda_{2}\ldots\lambda_{s-1}}\label{equToSolve}
\end{gather}
and put the traceless condition only in the end. Plugging the
expansions (\ref{f-exp}) and (\ref{X-exp}) into the above equation
leads to the conditions
\begin{gather}
f_{(n_{k})(n_{\bar{k}})i_{1}\ldots i_{s-1-n_{k}-n_{\bar{k}}}}  =  \tfrac{1}{k \bar{k}\vphantom{\big|}}X_{(n_{k})(n_{\bar{k}}-1)i_{1}\ldots i_{s-1-n_{k}-n_{\bar{k}}}}\qquad\forall \; n_{\bar{k}}\geq1 .\label{Xf-conds}
\end{gather}
The expansion coef\/f\/icients $f_{(n_{k})(0)i_{1}\ldots i_{s-1-n_{k}}}$
remain undetermined and can be used to make the solution traceless,
as we will see now. The trace of $f$ in (\ref{f-exp}) is given
by \begin{gather*}
f'_{\lambda_{1}\ldots\lambda_{s-3}}  =  \sum_{n_{k}=0}^{s-3}\sum_{n_{\bar{k}}=0}^{s-3-n_{k}}\left(2(k\bar{k})f_{(n_{k}+1)(n_{\bar{k}}+1)i_{1}\ldots i_{s-3-n_{k}-n_{\bar{k}}}}-\sum_{j=1}^{d-2}f_{(n_{k})(n_{\bar{k}})i_{1}\ldots i_{s-3-n_{k}-n_{\bar{k}}}jj}\right)  \\
\phantom{f'_{\lambda_{1}\ldots\lambda_{s-3}}  =}{}  \times\sum_{(n_{k})}\big(k_{\lambda_{1}}\cdots k_{\lambda_{n_{k}}}\big)\sum_{(n_{\bar{k}})}\big(\bar{k}_{\lambda_{n_{k}+1}}
\cdots\bar{k}_{\lambda_{n_{k}+n_{\bar{k}}}}\big)e_{\lambda_{n_{k}+n_{\bar{k}}+1}}^{i_{1}}\cdots e_{\lambda_{s-3}}^{i_{s-3-n_{k}-n_{\bar{k}}}} .
 \end{gather*}
For $f$ to be traceless, we have thus the additional condition
\begin{gather*}
\sum_{j=1}^{d-2}f_{(n_{k})(n_{\bar{k}})i_{1}\ldots i_{s-3-n_{k}-n_{\bar{k}}}jj}  =  2(k\bar{k})f_{(n_{k}+1)(n_{\bar{k}}+1)i_{1}\ldots i_{s-3-n_{k}-n_{\bar{k}}}}\\
\forall \; n_{k}\in\{0,\ldots,s-3\},\quad n_{\bar{k}}\in\{0,\ldots,s-3-n_{k}\}.
\end{gather*}
For $n_{\bar{k}}\geq1$ this becomes (using (\ref{Xf-conds}))
\begin{gather*}
\sum_{j=1}^{d-2}X_{(n_{k})(n_{\bar{k}}-1)i_{1}\ldots i_{s-3-n_{k}-n_{\bar{k}}}jj}  =  2(k\bar{k})X_{(n_{k}+1)(n_{\bar{k}})i_{1}\ldots i_{s-3-n_{k}-n_{\bar{k}}}} \\
\forall\; n_{k}\in\{0,\ldots,s-4\},\quad n_{\bar{k}}\in\{1,\ldots,s-3-n_{k}\}
\end{gather*}
and is automatically fulf\/illed due to the tracelessness
of $X$ (i.e.\ the double tracelessness of the current $J$). For
$n_{\bar{k}}=0$, however, we get the additional condition
\begin{gather}
\sum_{j=1}^{d-2}f_{(n_{k})(0)i_{1}\ldots i_{s-3-n_{k}}jj}  =  2X_{(n_{k}+1)(0)i_{1}\ldots i_{s-3-n_{k}}} \qquad \forall \; n_{k}\in\{0,\ldots,s-3\}.
\label{traceless-cond-on-f}\end{gather}
At this point it is useful to know the general form of the traceless-projector
$\Pi$ given in (\ref{tl-part}). 
Certainly, the projector obeys $\tr(A-\Pi A)=\tr A$
for any dimension and rank and the explicit form of the projector
shows that $(A-\Pi A)$ does not contain $A$ itself but is some function~$F$ of only $\tr A$ and higher traces, i.e.\ $(A-\Pi A)=F(\tr A)$.
One thus can replace~$\tr A$ in $\tr F(\tr A)=\tr A$ by an untraced
tensor $B$ and obtains $\tr F(B)=B$, so that the function $F$ provides
a solution for~(\ref{traceless-cond-on-f}). In our case we have the ef\/fective
dimension $d-2$ (instead of $d$) and the rank $s-1-n_{k}$ (instead
of $s$). The solution for the above equation~(\ref{traceless-cond-on-f})
is thus given by
\begin{gather}
f_{(n_{k})(0)i_{1}\ldots i_{s-1-n_{k}}}=
 -2\sum_{l=1}^{[(s-1-n_{k})/2]}\frac{1}{l!(-2)^{l}}\tfrac{(d/2+s-l-4-n_{k})!}{(d/2+s-4-n_{k})!}\label{cond-for-f-tl} \\
\phantom{f_{(n_{k})(0)i_{1}\ldots i_{s-1-n_{k}}}= }{}\times \sum_{2}\delta_{i_{1}i_{2}}\cdots\!\sum_{2}\delta_{i_{2l-1}i_{2l}}\!\sum_{j_{1}=1}^{d-2}\cdots\!\sum_{j_{l-1}=1}^{d-2}X_{(n_{k}+1)(0)j_{1}j_{1}\ldots j_{l-1}j_{l-1}i_{2l+1}\ldots i_{s-1-n_{k}}}\nn\\
\forall \;  n_{k}\in\{0,\ldots,s-3\}.\nonumber
\end{gather}
Now it remains to extract the expansion coef\/f\/icients $X_{(n_{k})(n_{\bar{k}}-1)i_{1}\ldots i_{s-1-n_{k}-n_{\bar{k}}}}$
from $X_{\lambda_{1}\ldots\lambda_{s-2}}$ via
\begin{gather}
X_{(n_{k})(n_{\bar{k}}-1)i_{1}\ldots i_{s-1-n_{k}-n_{\bar{k}}}}  =  \frac{1}{(k\bar{k})^{n_{k}+n_{\bar{k}}-1}}(\bar{k}^{\rho})^{n_{k}}(k^{\rho})^{n_{\bar{k}}-1}X_{\roro i_{1}\ldots i_{s-1-n_{k}-n_{\bar{k}}}}.\label{extr-exp-coeff}
\end{gather}
Plugging these coef\/f\/icients into (\ref{Xf-conds}) and this in turn
into the basis-expansion of $f$ yields the f\/inal solution of (\ref{equToSolve}) or (\ref{claim})
\begin{gather*}
f_{\lambda_1\ldots\lambda_{s-1}}= \sum_{n_{k}=0}^{s-2}\sum_{n_{\bar{k}}=1}^{s-1-n_{k}}\frac{1}{(k\bar{k})^{n_{k}+n_{\bar{k}}}}\bar{k}^{\rho_1}\cdots\bar k^{\rho_{n_k}}k^{\rho_{n_k+1}}\cdots k^{\rho_{n_k+n_{\bar k}-1}}X_{\rho_1\ldots\rho_{n_k+n_{\bar k}-1} i_{1}\ldots i_{s-1-n_{k}-n_{\bar{k}}}}
\nn\\
\phantom{f_{\lambda_1\ldots\lambda_{s-1}}= }{} \times\sum_{(n_{k})}\big(k_{\lambda_1}\cdots k_{\lambda_{n_k}}\big)\sum_{(n_{\bar{k}})}\big(\bar{k}_{\lambda_{n_k+1}\dots\lambda_{n_k+n_{\bar k}}}\big)\big(e_{\lambda_{n_k+n_{\bar k}+1}}^{i_{1}}\cdots e_{\lambda_{s-1}}^{i_{s-1-n_{k}-n_{\bar{k}}}}\big)\nonumber \\
\phantom{f_{\lambda_1\ldots\lambda_{s-1}}= }{} +\sum_{n_{k}=0}^{s-1}f_{(n_{k})(0)i_{1}\ldots i_{s-1-n_{k}}}\sum_{(n_{k})}\big(k_{\lambda_1}\cdots k_{\lambda_{n_k}}\big)\big(e_{\lambda_{n_k+1}}^{i_{1}}\cdots e_{\lambda_{s-1}}^{i_{s-1-n_{k}}}\big),
 \end{gather*}
where (due to (\ref{cond-for-f-tl}) and (\ref{extr-exp-coeff})) some coef\/f\/icients of the last
row are given by
\begin{gather*}
f_{(n_{k})(0)i_{1}\ldots i_{s-1-n_{k}}}=
-\frac{2}{(k\bar{k})^{n_{k}+1}}\bar{k}^{\rho_{1}}\cdots\bar{k}^{\rho_{n_{k}+1}}\sum_{l=1}^{[(s-1-n_{k})/2]}
\frac{1}{l!(-2)^{l}}\tfrac{(d/2+s-l-4-n_{k})!}{(d/2+s-4-n_{k})!}\nonumber \\
\phantom{f_{(n_{k})(0)i_{1}\ldots i_{s-1-n_{k}}}=}{} \times  \sum_{2}\delta_{i_{1}i_{2}}\ldots\!\sum_{2}\delta_{i_{2l-1}i_{2l}}\!\sum_{j_{1}=1}^{d-2}\ldots\!\sum_{j_{l-1}=1}^{d-2}
X_{\rho_{1}\ldots\rho_{n_{k}+1}j_{1}j_{1}\ldots j_{l-1}j_{l-1}i_{2l+1}\ldots i_{s-1-n_{k}}}\\
\forall \;   n_{k}\in\{0,\ldots,s-3\}.
\end{gather*}
Note that the coef\/f\/icients $f_{(s-2)(0)i_{1}\ldots i_{s-1-n_{k}}}$ and $f_{(s-1)(0)i_{1}\ldots i_{s-1-n_{k}}}$
are still undetermined and can be chosen arbitrarily.

For $s=3$ we have
\begin{gather*}
f_{(0)(0)i_{1}i_{2}}  =  \frac{2}{(d-2)(k\bar{k})}\bar{k}^{\rho}\delta_{i_{1}i_{2}}X_{\rho}
\end{gather*}
and the solution  becomes
\begin{gather*}
f_{\lambda_{1}\lambda_{2}}  =  \frac{1}{k\bar{k}}X_{i_{1}}\sum_{1}\bar{k}_{\lambda_{1}}e_{\lambda_{2}}^{i_{1}}
+\frac{1}{(k\bar{k})^{2}}k^{\rho}X_{\rho}\sum_{2}\bar{k}_{\lambda_{1}}\bar{k}_{\lambda_{2}}
+\frac{1}{(k\bar{k})^{2}}\bar{k}^{\rho}X_{\rho}\sum_{1}k_{\lambda_{1}}\bar{k}_{\lambda_{2}}\nonumber \\
\phantom{f_{\lambda_{1}\lambda_{2}}  =}{}  +\frac{2}{(d-2)k\bar{k}}\bar{k}^{\rho}X_{\rho}\sum_{i=1}^{d-2}e_{\lambda_{1}}^{i}e_{\lambda_{2}}^{i}
 +f_{(1)(0)i_{1}}\sum_{1}k_{\lambda_{1}}e_{\lambda_{2}}^{i_{1}}+f_{(2)(0)}k_{\lambda_{1}}k_{\lambda_{2}}.
 \end{gather*}
For an appropriate choice of $f_{(1)(0)i_{1}}$ and $f_{(2)(0)}$ this coincides in fact with the solution (\ref{fresultforsisthree}) given in the beginning, if
the metric $\eta^{\mu\nu}$ in (\ref{fresultforsisthree}) is expanded as in (\ref{onmassshell}).

\section{Symmetrization}
\label{app:symmetrization}

For the tensor product of two symmetric tensors it is very convenient
to introduce the symmetrized sums which run over  those permutations
of uncontracted indices which lead to inequivalent terms. For example
\begin{gather}
\sum_{1}A_{\lambda_{1}}B_{\lambda_{2}\ldots\lambda_{s}}  \equiv  \us{A_{\lambda_{1}}
B_{\lambda_{2}\ldots\lambda_{s}}+A_{\lambda_{2}}B_{\lambda_{1}\lambda_{3}\ldots\lambda_{s}}+\cdots+A_{\lambda_{s}}
B_{\lambda_{1}\ldots\lambda_{s-1}}}{s\:{\rm terms}},\label{sym-sum-one}\\
\sum_{2}A_{\lambda_{1}\lambda_{2}}B_{\lambda_{3}\ldots\lambda_{s}}  \equiv
\sum_{i<j}A_{\lambda_{i}\lambda_{j}}\us{B_{\lambda_{1}\ldots\lambda_{i-1}\lambda_{i+1}\ldots
\lambda_{j-1}\lambda_{j+1}\ldots\lambda_{s}}}{s(s-1)/2\:{\rm terms}}.\label{sym-sum-two}
\end{gather}
This dif\/fers from the projection to the symmetric part (denoted by a round bracket around the indices)
only by a normalization factor. In the above cases we have $A_{(\lambda_1}B_{\lambda_2\ldots\lambda_s)}=
\tfrac{1}{s}\sum_1 A_{\lambda_1}B_{\lambda_2\ldots\lambda_s}$ and $A_{(\lambda_1\lambda_2}
B_{\lambda_3\ldots\lambda_s)}=\tfrac{2}{s(s-1)}\sum_2 A_{\lambda_1\lambda_2}B_{\lambda_2\ldots\lambda_s}$.
In general the projection to the symmetric part is given by
\begin{gather*}
X_{(\lambda_{1}\ldots\lambda_{s})}  \equiv  \tfrac{1}{s!}
\sum_{\text{\rm all Perm's }P}X_{\lambda_{P(1)}\ldots\lambda_{P(s)}} .
\end{gather*}
It has the projection-property $X_{((\lambda_1\ldots\lambda_s))}=X_{(\lambda_1\ldots\lambda_s)}$ and appears
automatically when $X$ is contracted with any other symmetric tensor, in particular with the $s$-th power
of a vector $X_{\lambda_1\ldots\lambda_s}v^{\lambda_1}\cdots v^{\lambda_s}=X_{(\lambda_1\ldots\lambda_s)}
v^{\lambda_1}\cdots v^{\lambda_s}$.

The dif\/ferent normalization in the symmetrized sums (\ref{sym-sum-one}) and (\ref{sym-sum-two}) is in turn
more convenient in calculations. To study the general properties of theses symmetrized sums, let us f\/irst
note that one can extend the def\/inition to a product of a symmetric rank $p$ and a rank $q$ tensor.
It can be def\/ined as
\begin{gather}
\sum_{(p)}A_{\lambda_{1}\ldots\lambda_{p}}^{(p)}B_{\lambda_{p+1}\ldots\lambda_{p+q}}^{(q)}  \equiv
\tbinom{p+q}{p}A_{(\lambda_{1}\ldots\lambda_{p}}^{(p)}B_{\lambda_{p+1}\ldots\lambda_{p+q})}^{(q)}\nonumber \\
\phantom{\sum_{(p)}A_{\lambda_{1}\ldots\lambda_{p}}^{(p)}B_{\lambda_{p+1}\ldots\lambda_{p+q}}^{(q)}}{}
=  \tfrac{1}{p!q!} \sum_{\text{\rm all Perm's} \ P}  A_{\lambda_{P(1)}\ldots\lambda_{P(p)}}^{(p)}
 B_{\lambda_{P(p+1)}\ldots\lambda_{P(s)}}^{(q)}.\label{AsymB}
 \end{gather}
We have put the $p$ below the sum in brackets in order to avoid confusion with a sum over all~$p$.
Instead~$p$ is f\/ixed here. Again this symmetrized sum can be understood as the sum over all~$\tbinom{p+q}{p}$
inequivalent terms.
Because of $\tbinom{p+q}{p}\tbinom{p+q+r}{p+q}=
\tbinom{p+q+r}{p}\tbinom{q+r}{q}=\tfrac{(p+q+r)!}{p!q!r!}$,
this symmetrized sum is associative in the sense
\begin{gather*}
\sum_{(p+q)}\Bigl(\sum_{(p)}A^{(p)}B^{(q)}\Bigr)C^{(r)}  =
\sum_{(p)}A^{(p)}\Bigl(\sum_{(q)}
B^{(q)}C^{(r)}\Bigr) .
\end{gather*}
In addition it has a Leibniz-like behaviour with respect to index
contractions. If we denote the result of~(\ref{AsymB}) by
\begin{gather*}
C^{(p+q)}_{\lambda_1\ldots\lambda_{p+q}}\equiv\sum_{(p)}
A_{\lambda_{1}\ldots\lambda_{p}}^{(p)}B_{\lambda_{p+1}\ldots\lambda_{p+q}}^{(q)}
\end{gather*}
then the contraction with a vector acts like a derivative
\begin{gather*}
v^{\mu}C_{\mu\lambda_2\ldots\lambda_{p+q}}^{(p+q)}=
\sum_{(p-1)}v^{\mu}A_{\mu\lambda_2\ldots\lambda_{p}}^{(p)}
B_{\lambda_{p+1}\ldots\lambda_{p+q}}^{(q)}+\sum_{(p)}A_{\lambda_2\ldots\lambda_{p+1}}^{(p)}v^{\mu}
B_{\mu\lambda_{p+2}\ldots\lambda_{p+q}}^{(q)}  .
\end{gather*}
Contractions with rank $r$ tensors act like derivatives of order $r$
\begin{gather*}
D_{(r)}^{\mu_{1}\ldots\mu_{r}}C_{\mu_{1}\ldots\mu_{r}\lambda_{1}\ldots\lambda_{p+q-r}}^{(p+q)}
  =  \sum_{(p-r)}A_{\lambda_{1}\ldots\lambda_{p-r}\mu_{1}\ldots\mu_{r}}^{(p)}D_{(r)}^{\mu_{1}\ldots\mu_{r}}
 B_{\lambda_{p-r+1}\ldots\lambda_{p+q-r}}^{(q)}\nonumber \\
\phantom{D_{(r)}^{\mu_{1}\ldots\mu_{r}}C_{\mu_{1}\ldots\mu_{r}\lambda_{1}\ldots\lambda_{p+q-r}}^{(p+q)}=}{}
 +\sum_{(p-r+1)}rA_{\lambda_{1}\ldots\lambda_{p-r+1}\mu_{1}\ldots\mu_{r-1}}^{(p)}D_{(r)}^{\mu_{1}\ldots\mu_{r}}
 B_{\mu_{r}\lambda_{p-r+2}\ldots\lambda_{p+q-r}}^{(q)}  +\cdots{} \\
\phantom{D_{(r)}^{\mu_{1}\ldots\mu_{r}}C_{\mu_{1}\ldots\mu_{r}\lambda_{1}\ldots\lambda_{p+q-r}}^{(p+q)}=}{} +\sum_{(p-i)}\tbinom{r}{i}A_{\lambda_{1}\ldots\lambda_{p-i}\mu_{1}\ldots\mu_{i}}^{(p)}D_{(r)}^{\mu_{1}
 \ldots\mu_{r}}B_{\mu_{i+1}\ldots\mu_{r}\lambda_{p-i+1}\ldots\lambda_{p+q-r}}^{(q)}+\cdots{} \\
\phantom{D_{(r)}^{\mu_{1}\ldots\mu_{r}}C_{\mu_{1}\ldots\mu_{r}\lambda_{1}\ldots\lambda_{p+q-r}}^{(p+q)}=}{}
  +\sum_{(p)}A_{\lambda_{1}\ldots\lambda_{p}}^{(p)}D_{(r)}^{\mu_{1}\ldots\mu_{r}}
 B_{\mu_{1}\ldots\mu_{r}\lambda_{p+1}\ldots\lambda_{p+q-r}}^{(q)}.
 \end{gather*}
This behaviour is very convenient for calculating divergencies
(contracting with $k^\mu$) or
traces (contracting with the metric $\eta^{\mu\nu}$).

\section{Projection to the (double) traceless part}
\label{app:traceless}


For higher rank symmetric tensors the trace is def\/ined by contracting
any pair of indices with the metric and will be denoted by
\begin{gather*}
A'_{\lambda_{3}\ldots\lambda_{s}}\equiv\eta^{\lambda_{1}\lambda_{2}}
A_{\lambda_{1}\lambda_{2}\lambda_{3}\ldots\lambda_{s}}.
\end{gather*}
The projector $\proj_{\lambda_1\ldots\lambda_s}\hoch{\rho_1\ldots\rho_s}$
to the {\it traceless part} of an arbitrary symmetric rank $s$ tensor f\/ield
$A_{\lambda_{1}\ldots\lambda_{s}}$ is then given by
\begin{gather}
(\Pi A)_{\lambda_{1}\ldots\lambda_{s}}  \equiv  A_{\lambda_{1}\ldots\lambda_{s}}-
\tfrac{1}{d+2s-4}\sum_{2}\eta_{\lambda_{1}\lambda_{2}}A'_{\lambda_{3}\ldots\lambda_{s}}\nonumber \\
\phantom{(\Pi A)_{\lambda_{1}\ldots\lambda_{s}}  \equiv}{} +\sum_{k=2}^{[s/2]}\tfrac{1}{k!(-2)^{k}}\tfrac{(d/2+s-k-2)!}{(d/2+s-2)!}
 \underbrace{\sum_{2}\eta_{\lambda_{1}\lambda_{2}}\ldots\sum_{2}\eta_{\lambda_{2k-1}\lambda_{2k}}
 A_{\lambda_{2k+1}\ldots\lambda_{s}}^{('k)}}_{\tfrac{s!}{2^{k}(s-2k)!}\mbox{ terms}},
\label{tl-part}
 \end{gather}
where $A^{('k)}$ is the $k$-th trace of $A$, $[s/2]\equiv\left\{\zwek{s/2\mbox{ for }s
\mbox{ even}}{(s-1)/2\mbox{ for }s\mbox{ }\mbox{odd}}\right.$
is the integer part of $s/2$ and the factorials have to be understood
via the $\Gamma$-function in the case of an odd dimension~$d$. One can convince oneself
that $(\Pi A)'=0$ for any $A$. When $A$ is double traceless ($A''=0$), only the f\/irst
line contributes to the projection. When $A$ is already traceless, the projector reduces
to the identity as it should. One could def\/ine various dif\/ferent projections to the
subspace of traceless tensors. The important property of this one is the fact that the
operator is symmetric in the sense $\proj_{\lambda_{1}\ldots\lambda_{s}}\hoch{\rho_{1}
\ldots\rho_{s}}=\proj\hoch{\rho_{1}\ldots\rho_{s}}\tief{\lambda_{1}\ldots\lambda_{s}}$.
The existence of a symmetric projection operator to some subspace always guarantees that
tensors contracted with some subspace element get projected to the same subspace.
If we contract for example the traceless tensor  $(\Pi A)_{\lambda_1\ldots\lambda_s}$
with an arbitray tensor $B_{\lambda_1\ldots\lambda_s}$, we can write
$(\Pi A)_{\lambda_1\ldots\lambda_s}B^{\lambda_1\ldots\lambda_s}=A_{\lambda_1\ldots\lambda_s}
(\Pi B)^{\lambda_1\ldots\lambda_s}$.

Similarly to above, one can project to a {\it double traceless part}, where only double
and higher traces are subtracted appropriately
\begin{gather*}
(\proj^{(dtl)}A)_{\lambda_{1}\ldots\lambda_{s}}  \equiv  A_{\lambda_{1}\ldots\lambda_{s}}
 -\sum_{k=2}^{[s/2]}\tfrac{k-1}{(-2)^{k}k!}\cdot\tfrac{(d/2+s-k-3)!}{(d/2+s-3)!}
 \underbrace{\sum_{2}\eta_{\lambda_{1}\lambda_{2}}\ldots\sum_{2}\eta_{\lambda_{2k-1}\lambda_{2k}}
 A_{\lambda_{2k+1}\ldots\lambda_{s}}^{('k)}}_{\tfrac{s!}{2^{k}(s-2k)!}\mbox{ terms}}.
 \end{gather*}
Again one can check that indeed $(\proj^{(dtl)}A)''=0$ for all symmetric rank $s$ tensor f\/ields $A$.
We will not make use of this projection operator in the main text, but it is important to keep in
mind that this  projector would have been needed, if the variation of the Fronsdal
action~(\ref{FronsdalAction}) with respect to the (double traceless) tensor gauge f\/ield did not
automatically produce a double traceless expression~(\ref{LA1}). Note f\/inally that also this
projector acts symmetrically. This is the reason why the double traceless property of the
current $J$ is inherited by the contraction with the double traceless tensor gauge f\/ield~$A$
in the action~(\ref{FronsdalAction}).

\subsection*{Acknowledgements}
The work of S.G. was supported by ENRAGE (European Network on Random
Geometry), a~Marie Curie Research Training Network, contract MRTN-CT-2004-005616. The work of G.S. was partially supported by the EEC Grant
no. MRTN-CT-2004-005616.

\pdfbookmark[1]{References}{ref}
\LastPageEnding

\end{document}